\documentclass[10pt,aps,prd,preprintnumbers,showpacs,twocolumn,nofootinbib,notitlepage,longbibliography,superscriptaddress]{revtex4-1}

\usepackage{amsmath, amssymb,braket}
\usepackage[dvipsnames]{xcolor}
\usepackage{color}
\usepackage{float}
\usepackage{graphicx}
\usepackage{subfigure}
\usepackage{natbib}
\usepackage{tikz}
\usetikzlibrary{shapes,arrows}
\usepackage[compat=1.1.0]{tikz-feynman}
\usepackage[colorlinks=true
,urlcolor=blue
,anchorcolor=blue
,citecolor=blue
,filecolor=blue
,linkcolor=red
,menucolor=blue
,hyperfootnotes=false,
,linktocpage=true
,pdfproducer=medialab
,pdfa=true
]{hyperref}
\usepackage{siunitx}
\usepackage{multirow}

\DeclareSIUnit \h {\ensuremath{\mathit{h}}}
\DeclareSIUnit\eV{e\kern-.05em V}
\DeclareSIUnit\parsec{pc}
\DeclareSIUnit\year{yr}
\DeclareSIUnit\erg{erg}
\DeclareSIUnit\sr{sr}

\def\jcap{\rm{J. Cosmology Astropart. Phys.}}
\begin{document}
\title{CMB limits on decaying dark matter beyond the ionization threshold}

\preprint{MIT-CTP/5747}

\author{Clara Xu}
\email{xuclara@berkeley.edu}
\affiliation{Center for Theoretical Physics, Massachusetts Institute of Technology, Cambridge, Massachusetts 02139, USA}
\affiliation{Department of Physics, University of California, Berkeley, California 94720, USA}

\author{Wenzer Qin}
\email{wenzer.qin@nyu.edu}
\affiliation{Center for Theoretical Physics, Massachusetts Institute of Technology, Cambridge, Massachusetts 02139, USA}
\affiliation{Center for Cosmology and Particle Physics, Department of Physics, New York University, New York, NY 10003, USA}

\author{Tracy~R. Slatyer}
\email{tslatyer@mit.edu}
\affiliation{Center for Theoretical Physics, Massachusetts Institute of Technology, Cambridge, Massachusetts 02139, USA}

\begin{abstract}
    The temperature and polarization anisotropies of the cosmic microwave background (CMB) have been used to set constraints on decaying dark matter models down to keV masses. 
    In this work, we extend these limits down to the sub-keV mass range. 
    Using principal component analysis, we estimate the lower bound on the decay lifetime for a basis of different dark matter masses and Standard Model final states, from which the bound on an arbitrary model can be calculated. 
    We validate our principal component analysis using Markov chain Monte Carlo methods and \textit{Planck} 2018 data.
    We perform a separate analysis for models decaying into photons below the hydrogen ionization threshold. 
    We demonstrate that for these models, the effect of energy deposition can be captured approximately by a single parameter, but the redshift dependence of the effect is very different from higher-energy injections; in particular, the perturbations to CMB anisotropies are more sensitive to energy deposited around the time of recombination. 
\end{abstract}

\maketitle

%%%%%%%%%%%%%%%%%%%%%%%%%%%%%%%%%%%%%%%%%%%%%%%%%%%%%%%%%%%%%%%%%
%%%%%%%%%%%%%%%%%%%%%%%%%%%%%%%%%%%%%%%%%%%%%%%%%%%%%%%%%%%%%%%%%
\section{Introduction}
\label{sec:intro}
%%%%%%%%%%%%%%%%%%%%%%%%%%%%%%%%%%%%%%%%%%%%%%%%%%%%%%%%%%%%%%%%%
%%%%%%%%%%%%%%%%%%%%%%%%%%%%%%%%%%%%%%%%%%%%%%%%%%%%%%%%%%%%%%%%%

The fundamental nature of dark matter (DM) has been an open question since the establishment of modern physical cosmology.
The fact that the DM and baryonic energy densities differ by only an order one factor hint at possible interactions relating the abundances of the two sectors, although these interactions may be very weak today.
Interactions such as DM decays or annihilations into Standard Model (SM) particles may continue to inject energy into the visible universe in the form of e.g. heat or ionization; we can search for imprints of such injections in astrophysical data.

For example, measurements of the Lyman-$\alpha$ forest~\cite{Hiss:2017qyw, Walther:2018pnn, Gaikwad:2020art, Gaikwad:2020eip} have been used to constrain excess heating of the intergalactic medium (IGM) from decaying and annihilating dark matter~\cite{Cirelli:2009bb, Diamanti:2013bia, Liu:2016cnk, Liu:2020wqz, Capozzi:2023xie}, and dark photon dark matter~\cite{McDermott:2019lch, Caputo:2020bdy, Witte:2020rvb, Bolton:2022hpt}.
The effects of exotic heating on signals from 21\,cm cosmology have been used to forecast limits on energy injection from decaying and annihilating dark matter, as well as primordial black holes~\cite{Evoli:2014pva, Lopez-Honorez:2016sur, DAmico:2018sxd, Liu:2018uzy, Cheung:2018vww, Mitridate:2018iag, Clark:2018ghm, 2022JCAP...03..030M}.
Exotic injections can impact radiation backgrounds such as the CMB blackbody spectrum~\cite{Liu:2023fgu,Liu:2023nct}, which has also been used to constrain dark matter energy injection~\cite{Chluba:2013vsa,Ali-Haimoud:2015pwa,Chluba:2015hma,Acharya:2018iwh}.
Finally, a number of groups have studied the impact of exotic energy injection on the formation of the earliest compact objects in the universe~\cite{Ripamonti:2006gr,Pandey:2018jun,Friedlander:2022ovf,Qin:2023kkk,Lu:2023xoi,Lu:2024zwa}.

The anisotropies of the Cosmic Microwave Background (CMB) are a particularly robust probe of exotic energy injection, since the CMB anisotropies are well-understood and can be predicted by linear cosmological perturbation theory. Furthermore, it has been shown that the signal is mostly controlled by the amount of energy injected in electromagnetic particles, and is not particularly sensitive to the spectrum of injected particles (provided that the energy injected is above the hydrogen ionization threshold), allowing for constraints on a broad range of models.
Searches for effects on the power spectrum have been used to set constraints on both dark matter annihilation and decay~\cite{Adams:1998nr, Chen:2003gz, Padmanabhan:2005es, Slatyer:2009yq, Kanzaki:2009hf, Slatyer:2015jla, Slatyer:2016qyl, Poulin:2016anj, Cang:2020exa, Capozzi:2023xie}.
However, most of these limits only extend down to a DM mass of 10 keV, since the methods for calculating the cooling cascades of injected particles used approximations that were not validated below this injection energy.

The code package \texttt{DarkHistory} calculates the global temperature and ionization history of the universe given an exotic source of energy injection, such as dark matter annihilation or decay~\cite{Liu:2019bbm}. 
With the \texttt{DarkHistory v2.0} upgrade, the code's treatment of energy deposition by low-energy electrons was made more general, used updated cross-sections, and was validated against other deposition codes~\cite{Liu:2023fgu,Liu:2023nct}.
In addition, the treatment of hydrogen was upgraded from a three-level atom to a multi-level atom, allowing the code to track the contributions from many more atomic transitions. Previously, in Refs.~\cite{Liu:2023fgu,Liu:2023nct}, the CMB bounds from \texttt{DarkHistory v2.0} were estimated based on the change to the ionization history at $z=300$, whereas Ref.~\cite{Capozzi:2023xie} performed a more complete calculation for the CMB bounds using \texttt{DarkHistory v1.0}, focusing on DM decay into energies above the hydrogen ionization threshold (as well as some limits for injection energies below 13.6 eV, but these results rely on the approximate treatment of excitation in \texttt{DarkHistory v1.0}). 

In this paper, we recalculate the constraints on exotic energy injection from the CMB power spectrum using \texttt{DarkHistory v2.0}. 
We present these results for DM decaying to photons and electron-positron pairs.
We also extend the photon constraints to energies below the hydrogen ionization threshold of 13.6 eV, as photons injected at energies lower than this are still able to ionize existing \textit{excited} atoms, or excite a neutral atom that is subsequently ionized by a CMB photon. 
\texttt{DarkHistory v2.0} allows us to precisely determine the modification to the ionization history due to such multi-step ionizations.

In Section~\ref{sec:eng_inj}, we describe how we calculate the effect of exotic energy injection on the CMB anisotropies, employing in particular the \texttt{DarkHistory} and \texttt{ExoCLASS} codes.
Next, in Section~\ref{sec:pca}, we perform a principal component and MCMC analysis to estimate and better understand the impact of decaying DM on the CMB power spectrum.
We summarize and compare our new constraints to existing ones in Section~\ref{sec:constraints}.
We conclude in Section~\ref{sec:conclusion}.
Throughout the text, we will use natural units, where $\hbar = c = 1$.

%%%%%%%%%%%%%%%%%%%%%%%%%%%%%%%%%%%%%%%%%%%%%%%%%%%%%%%%%%%%%%%%%
%%%%%%%%%%%%%%%%%%%%%%%%%%%%%%%%%%%%%%%%%%%%%%%%%%%%%%%%%%%%%%%%%
\section{Energy Injection}
\label{sec:eng_inj}
%%%%%%%%%%%%%%%%%%%%%%%%%%%%%%%%%%%%%%%%%%%%%%%%%%%%%%%%%%%%%%%%%
%%%%%%%%%%%%%%%%%%%%%%%%%%%%%%%%%%%%%%%%%%%%%%%%%%%%%%%%%%%%%%%%%

Decaying DM injects energy into the universe at the following rate:
\begin{equation}
    \begin{aligned}
        % \left(\frac{dE}{dtdV}\right)^\text{ann}_\text{inj} & = \frac{\braket{\sigma v}}{M_\chi} f_\chi^2 \Omega_\text{DM}^2 \rho_c^2 (1+z)^6 , \\
        \left(\frac{dE}{dtdV}\right)^\text{dec}_\text{inj} & = \frac{e^{-t/\tau}}{\tau} f_\chi \Omega_\text{DM} \rho_c (1+z)^3 ,
    \end{aligned}
\end{equation}
where 
%$\braket{\sigma v}$ is the thermally averaged cross section, 
$f_\chi$ is the fraction of DM that can decay, $\Omega_\text{DM}$ is the DM density parameter, $\rho_c$ is the critical density of the universe, $\tau$ is the DM decay lifetime, and $1+z$ is the redshift.

Understanding the effects of energy injection also requires specifying the injected spectrum of particles. 
DM can decay into both stable and unstable SM particles, but unstable particles decay quickly into stable particles, so we can limit ourselves to studying the effects of deposited stable particles. 
We further neglect energy deposited by neutrinos and protons/antiprotons, as neutrinos simply escape and the effect of protons/antiprotons is expected to be small~\cite{Weniger:2013hja}. Additionally, in the lower mass range of greatest interest to us in this paper, there is not enough energy to produce protons/antiprotons.
We therefore focus on DM decaying to pairs of photons or $e^+ e^-$, as these particles dominate the effect on the CMB. 

The injection of high-energy photons and electrons/positrons initiates an electromagnetic cascade through interactions with thermal photons and the baryonic gas, increasing the number of non-thermal particles while decreasing their average energy. For photons with energies from MeV up to GeV scales, the dominant processes are Compton scattering with background electrons and pair production on the gas; at higher energies, photon-photon scattering and pair production on the CMB dominate, whereas at lower energies, photoionization becomes efficient. Electrons/positrons dominantly lose energy through inverse Compton scattering at high energies ($\gg$ keV), but interactions with the gas come to dominate once they reach energies at keV levels.
Eventually, their energy is deposited into a number of different channels, including ionization or excitation of hydrogen and helium, and heating of the gas. 
Describing the energy deposition process accurately requires following the daughter particle spectra for large energy ranges and time scales, for which we use the \texttt{DarkHistory} code~\cite{Liu:2019bbm,Liu:2023fgu,Liu:2023nct}. The code employs transfer functions that take in the injected energy spectrum and output the resulting spectrum of low energy secondary particles.
The process is slightly different for DM decaying into photons below the ionization threshold at 13.6 eV, so we describe both processes in the following subsections.

%%%%%%%%%%%%%%%%%%%%%%%%%%%%%%%%%%%%%%%%%%%%%%%%%%%%%%%%%%%%%%%%%
\subsection{Decay into electrons and photons above ionization threshold}
%%%%%%%%%%%%%%%%%%%%%%%%%%%%%%%%%%%%%%%%%%%%%%%%%%%%%%%%%%%%%%%%%

\begin{figure}
    \centering
    \includegraphics[width=\columnwidth]{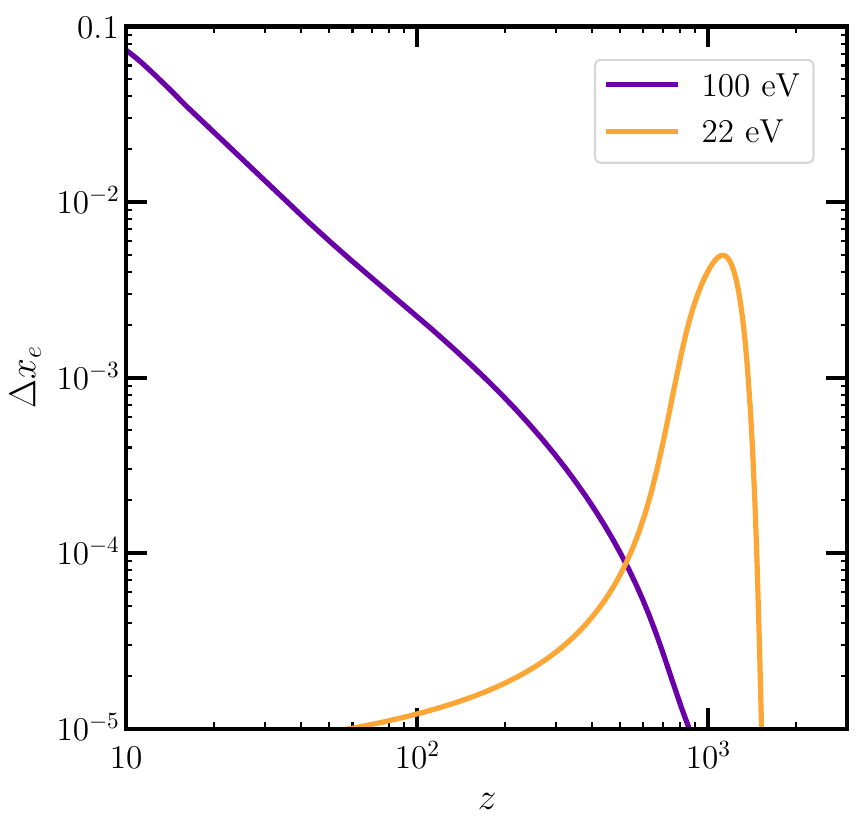}
    \caption{Absolute difference in the free electron fraction between the standard ionization history with no energy injection and the ionization history resulting from two energy injection models. One DM model decays into photon pairs with a mass of 100 eV and lifetime $2\times 10^{25}$ s. The other decays into photons with a mass of 22 eV and lifetime $5 \times 10^{21}$ s. The two models shown both have a lifetime that is a factor of few below the constraint on the lifetime from our PCA analysis, and are thus excluded by about the same amount.
    }
    \label{fig:delta_xe_highlowmass}
\end{figure}

As the primary injected particles propagate and cool, they may deposit their energy via a number of different channels.
The fraction of energy deposited into each channel can be tracked using \textit{energy deposition functions} $f_c(z)$ defined as 
\begin{equation}
    \left(\frac{dE}{dtdV}\right)_\text{dep,c} = f_c(z) \left(\frac{dE}{dtdV}\right)_\text{inj},
\end{equation}
where $c$ specifies the channel the energy is deposited in, e.g.~ionization or heating. These functions determine the effect of energy injection on the CMB.

Previously, the CMB anisotropies were used to set constraints on DM energy injection for $m_\chi \geq 10$ keV~\cite{Adams:1998nr, Chen:2003gz, Padmanabhan:2005es, Slatyer:2009yq, Kanzaki:2009hf, Slatyer:2015jla, Slatyer:2016qyl, Poulin:2016anj, Cang:2020exa}, where $m_\chi$ denotes the DM mass.
Ref.~\cite{Capozzi:2023xie} extended the constraints for decaying DM down to masses of 20.4 eV using a modified version of \texttt{DarkHistory v1.0} that allows for collisional excitation processes at all redshifts, and approximating $f_c(z)$ by its value at $z=300$ rather than the full function. 

For calculating the energy deposition functions, we use \texttt{DarkHistory v2.0}~\cite{Liu:2023fgu,Liu:2023nct}. 
\texttt{DarkHistory v2.0} includes an improved treatment of energy deposition by low-energy particles and a multi-level atom approximation for hydrogen that allows the code to track an arbitrary number of excited states of hydrogen.
These improvements allow for a more accurate calculation of the $f_c$'s at sub-keV injections. 
For all of the following results, we track up to $n=200$ hydrogen levels
\footnote{Using the iterative method described in Ref.~\cite{Liu:2023fgu} to calculate the recombination and ionization rates, we run five iterations to ensure these quantities are also converged.}
to ensure that the spectrum of low-energy photons is converged.

While energy injection can significantly increase the temperature of the intergalactic medium $T_b$, from the perspective of CMB anisotropies, the main impact of exotic energy deposition is to raise the global free electron fraction $x_e$.
The excess free electrons scatter CMB photons and thus modify the temperature and polarization  anisotropies, with a different redshift dependence than modifications to reionization.

To calculate the evolution of $x_e$ and $T_b$ with exotic energy injection, we use the \texttt{HyRec} code~\cite{Ali-Haimoud:2010hou} implemented in the \texttt{ExoCLASS} branch~\cite{Stocker:2018avm} of the \texttt{CLASS} code, with $f_c(z)$ functions from \texttt{DarkHistory} as inputs.
As an example, Fig.~\ref{fig:delta_xe_highlowmass} shows the change in the free electron fraction caused by DM with a mass of 100 eV decaying into photons with lifetime $2\times 10^{25}$ s. From the analysis in Ref.~\cite{Finkbeiner:2011dx}, we expect ionizations from $z \sim 300$ to dominate the change to the CMB anisotropies for decaying DM, at least for injection energies above the hydrogen ionization threshold. 

\begin{figure}
    \centering
    \includegraphics[width=\columnwidth]{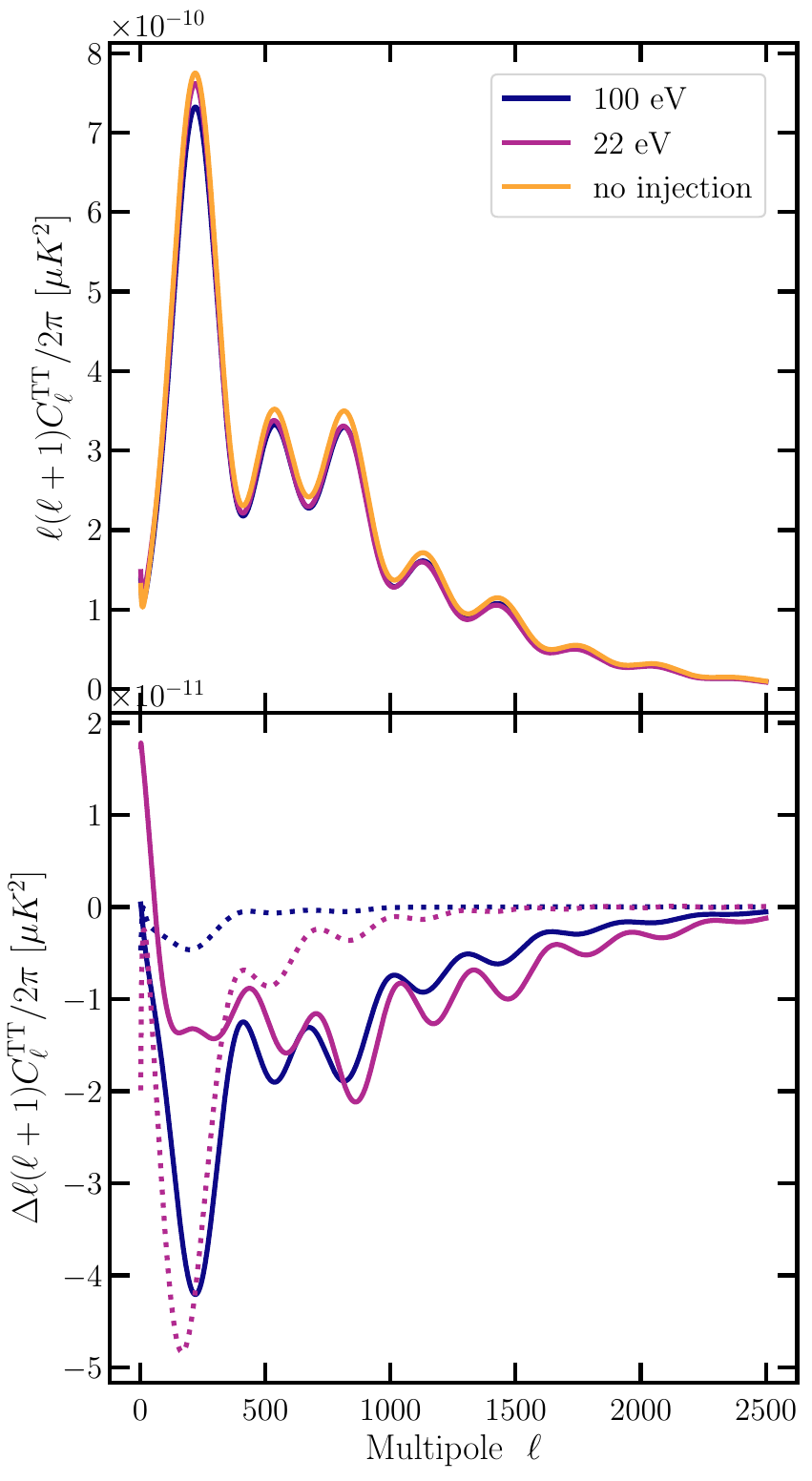}
    \caption{Comparison of the CMB $TT$ anisotropies for a model with no energy injection and the same two models including energy injection used in Fig.~\ref{fig:delta_xe_highlowmass}: one decaying into pairs of photons at a mass of 100 eV and lifetime $2\times 10^{25}$ s, one decaying to photons at a mass of 22 eV with lifetime $5 \times 10^{21}$ s. The bottom panel shows the change in $C_\ell^{TT}$ before (solid lines) and after (dotted lines) marginalizing over cosmological parameters.}
    \label{fig:mid_low_CLs}
\end{figure}

The main effect on the $C_\ell$'s is an overall damping of the peaks resulting from the increased optical depth at a large redshift range. The effect of Silk damping, important at higher multipoles, also increases due to the broadening of the surface of last scattering; these effects are degenerate with the slope and amplitude of the primordial power spectrum~\cite{Padmanabhan:2005es}. As an example, the $C_\ell^{TT}$'s resulting from the same energy injection model as above are shown in Fig.~\ref{fig:mid_low_CLs}, along with the change in $C_\ell^{TT}$ before and after marginalizing over the cosmological parameters listed in Section~\ref{sec:pca}.

%%%%%%%%%%%%%%%%%%%%%%%%%%%%%%%%%%%%%%%%%%%%%%%%%%%%%%%%%%%%%%%%%
\subsection{Decay into photons below ionization threshold}
%%%%%%%%%%%%%%%%%%%%%%%%%%%%%%%%%%%%%%%%%%%%%%%%%%%%%%%%%%%%%%%%%

\begin{figure*}
    \centering
    \includegraphics[width=\textwidth]{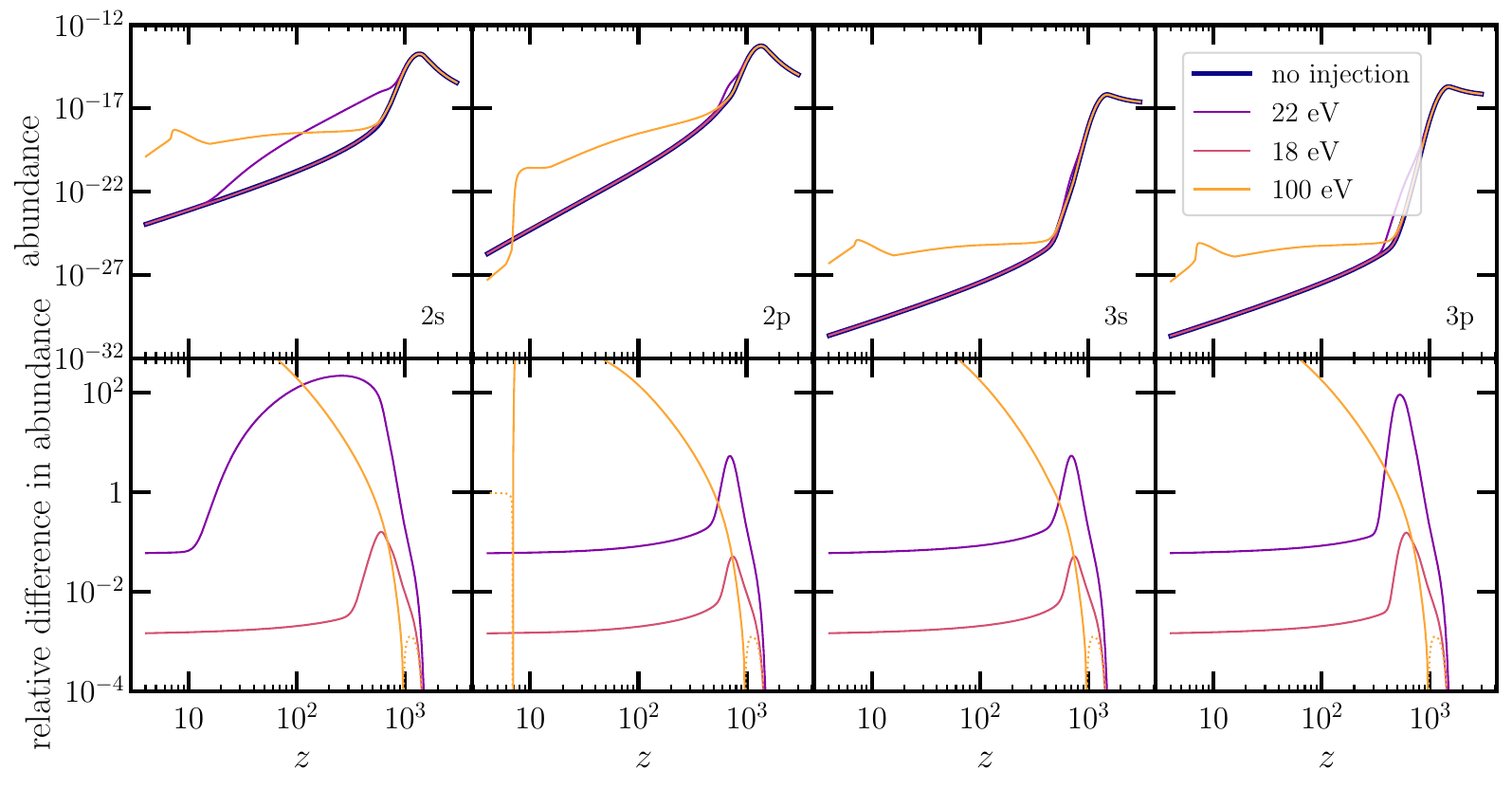}
    \caption{The panels in the top row show the fraction of atoms in the 2s, 2p, 3s, and 3p excited states of hydrogen. Each column represents one excited state. The bottom row shows the relative difference in the fraction compared to the case of no energy injection. Three energy injection models are shown. The first decays into photons with a DM mass of 22 eV with lifetime $5 \times 10^{21}$ s into two photons with total energy 22 eV; the second decays with the same lifetime but at a mass of 18 eV. The third decays with lifetime $2\times 10^{25}$ s at a mass of 100 eV. Dotted lines show the negative of the plotted values.}
    \label{fig:excstateabundances}
\end{figure*}

\begin{figure}
    \centering
    \includegraphics[width=\columnwidth]{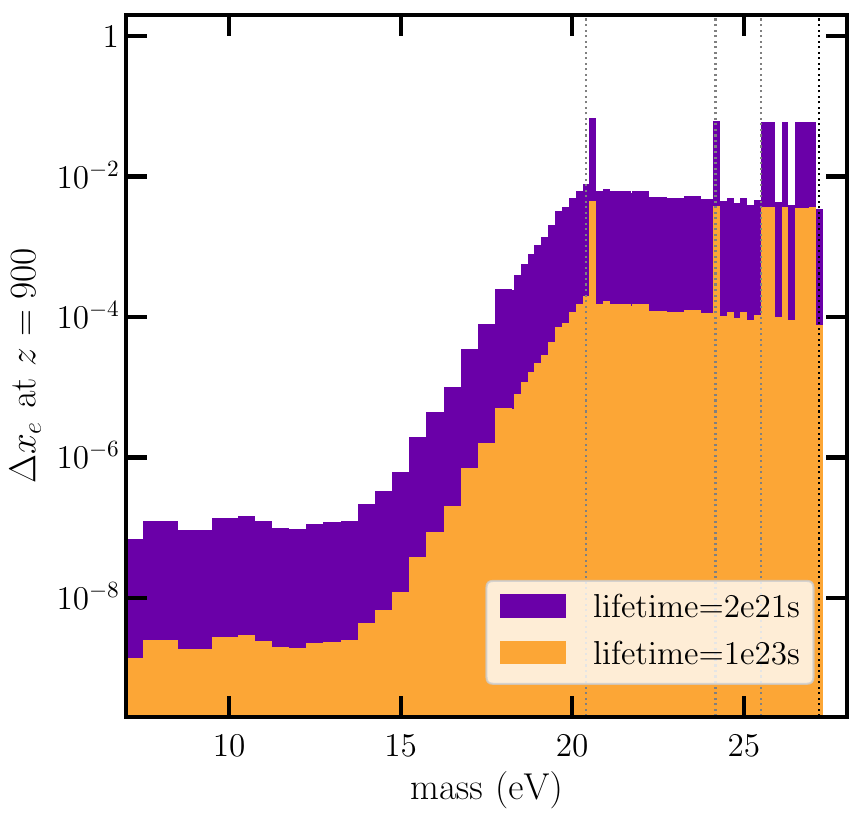}
    \caption{The absolute difference in the free electron fraction between the case with no energy injection and energy injection models at two lifetimes, $2 \times 10^{21}$ s and $1\times 10^{23}$ s, at $z=900$ around recombination. We plot this value over the DM mass; the DM will decay into two photons with the same total energy. The three vertical gray dotted lines represent twice the energy difference between hydrogen energy levels $n=1$ and $n=2,3,4$ respectively, and the black dotted line is at twice the ionization energy.}
    \label{fig:delta_xe_lowmass}
\end{figure}

For DM decaying into photons with energy less than $13.6$ eV, it is evident that the primary photons are unable to directly ionize hydrogen. 
However, between 10.2 and 13.6 eV, the injected photons are still  energetic enough to excite hydrogen from its ground state, allowing atoms to be ionized by background photons with energies of 3.4 eV or less (as the initial excitation may be to an excited state), which are exponentially more abundant than 13.6 eV photons at the time of recombination and later. Thus the presence of such excitation-inducing photons can still modify the $x_e$ fraction, especially at high redshifts where few-eV photons remain relatively abundant. 
At even lower energies, the injected photons could ionize existing excited atoms, although the effect on the free electron fraction will quickly diminish at lower redshifts as the abundance of higher excited states becomes very suppressed.
\texttt{ExoCLASS} only tracks excitations to the first excited state, and this is insufficient to accurately calculate the constraints below the ionization threshold. We therefore cannot rely on the energy deposition functions in this case. However, the energy deposition functions are only a means to calculate the ionization and temperature histories, which \texttt{DarkHistory} can compute accurately at these lower energies. We can therefore substitute the histories calculated in \texttt{DarkHistory} into \texttt{ExoCLASS} to use to compute the CMB anisotropies.

To fully realize the potential of  \texttt{DarkHistory v2.0} to treat photons at these energies, we must modify the code slightly. 
The transfer functions used to calculate daughter particle spectra from photon cooling must cover an energy range spanning several orders of magnitude, from CMB energies up to TeV scales. The default energy binning is therefore relatively coarse and cannot resolve the atomic lines at the lowest energies.
Fortunately, the lowest energy process that is included in the transfer functions is photoionization; that is, at initial energies below 13.6 eV, the transfer functions are merely given by the identity and this step of the evolution of the photon spectra becomes trivial.
Hence, we can bypass these transfer functions entirely.

For injections of photons below the ionization threshold, we modify \texttt{DarkHistory} such that the photons are not passed through the transfer functions (and hence are not rebinned) and are instead given directly to the atomic physics module, where we use a finer energy spacing consisting of two thousand logarithmically spaced bins between $10^{-8}$ and 13.6 eV.
We have tested that the binning is fine enough for the atomic line spectra and ionization history results to be converged.
We then modify \texttt{ExoCLASS} such that prior to reionization, instead of using \texttt{ExoCLASS}'s own calculated values for the thermal history, it takes the ionization fraction and matter temperature calculated using \texttt{DarkHistory}. We verified that for decay into electrons and photons above the ionization threshold, this method gives the same perturbations to CMB anisotropies as running \texttt{ExoCLASS} with $f_c(z)$ functions as inputs.

In their analysis of the mass range between 20.4 and 27.2 eV, Ref.~\cite{Capozzi:2023xie} assumed the ``excitation" channel $f_\mathrm{exc}$ as defined in \texttt{DarkHistory} captures the full effect of the photons on the ionization history. This channel integrates the photon power between 10.2 and 13.6 eV and assumes it all goes into excitation, whereas our analysis treats the full spectrum of photons and different lines of hydrogen, and so should be viewed as a more accurate update to the earlier analysis.

We find that for injection of photons below 13.6 eV, the effect on the free electron fraction is strongest during redshifts around recombination; Fig.~\ref{fig:delta_xe_highlowmass} shows a clear peak in $\Delta x_e$ at $z \sim 1000$ for DM of mass 22 eV decaying with a lifetime of $5 \times 10^{21}$ s.
Afterwards, as redshift decreases, the change in the free electron fraction $\Delta x_e$ quickly dies off. 
This is in line with the decreasing number density of CMB photons with sufficient energy to ionize atoms in excited states; as the CMB redshifts to lower energies, the abundance of $\sim 3$ eV photons in the blackbody tail decreases. 
In Fig.~\ref{fig:excstateabundances}, we plot the fraction of atoms in excited states 2s, 2p, 3s, and 3p and their relative difference with respect to a model with no energy injection. 
DM models with mass below the ionization threshold show a large increase in excited state abundances around recombination at $z \sim 1000$. 
There is an especially large increase in the 2s excited state for DM with mass 22 eV, which decays into two photons of 11 eV and thus produces a large amount of photons around the Lyman-$\alpha$ line able to contribute to two-photon excitations to the 2s state (together with one of the abundant lower-energy CMB photons).

In Fig.~\ref{fig:delta_xe_lowmass}, we plot the difference in $x_e$ at $z = 900$ for a range of masses below 27.2 eV. There are resonances at masses corresponding to spectral lines of hydrogen, further corroborating the importance of excited states in the effect on the ionization history.

We plot $C_\ell^{TT}$ as a function of $\ell$ for DM of mass 22 eV decaying into photons with a lifetime of $5 \times 10^{21}$ s in Fig.~\ref{fig:mid_low_CLs}, along with $\Delta C_\ell^{TT}$ before and after marginalizing over cosmological parameters. 
For models that decay into photons with energies above the ionization threshold, the main effect on the TT $C_\ell$'s before marginalization is a damping of the peaks. 
This is also true for these models, decaying into photons below 13.6 eV, with the exception that the \textit{first} peak is not damped to the same degree. 
This is possibly due to the effect of these models on $\Delta x_e$ being concentrated at earlier redshifts, meaning they have less effect on lower $\ell$s which are more sensitive to later redshifts. Interestingly, the change to $C_\ell^{TT}$ for these models is also less degenerate with cosmological parameters. 

Figs.~\ref{fig:delta_xe_highlowmass} and \ref{fig:excstateabundances} point to a difference in how models decaying into photons below the ionization threshold affect $x_e$ and thus CMB anisotropies. Compared to models of decay into electrons or photons above the ionization threshold, the effect on $x_e$ and excited state abundances is concentrated around recombination redshifts rather than later redshifts. The previous intuition of changes to the ionization level around $z \sim 300$ being the most important redshift for CMB anisotropies, for decaying DM, therefore does not hold here. 
We expect redshifts around $z \sim 1000$ to have a much greater impact, as will be confirmed in the next section.

%%%%%%%%%%%%%%%%%%%%%%%%%%%%%%%%%%%%%%%%%%%%%%%%%%%%%%%%%%%%%%%%%
%%%%%%%%%%%%%%%%%%%%%%%%%%%%%%%%%%%%%%%%%%%%%%%%%%%%%%%%%%%%%%%%%
\section{Principal Component Analysis}
\label{sec:pca}
%%%%%%%%%%%%%%%%%%%%%%%%%%%%%%%%%%%%%%%%%%%%%%%%%%%%%%%%%%%%%%%%%
%%%%%%%%%%%%%%%%%%%%%%%%%%%%%%%%%%%%%%%%%%%%%%%%%%%%%%%%%%%%%%%%%

%%%%%%%%%%%%%%%%%%%%%%%%%%%%%%%%%%%%%%%%%%%%%%%%%%%%%%%%%%%%%%%%%
\subsection{Methods}
%%%%%%%%%%%%%%%%%%%%%%%%%%%%%%%%%%%%%%%%%%%%%%%%%%%%%%%%%%%%%%%%%

Since the effects of energy injection from different DM models on CMB anisotropies are highly correlated, they can be characterized by a small number of parameters. Principal component analysis (PCA) provides a method of finding these parameters, giving a basis of principal components with orthogonal effects on CMB anisotropies. Energy injection models can be decomposed into this basis. As we will see later, the first term in this decomposition generally dominates the effect on the CMB.

We follow the methods described in Refs.~\cite{Slatyer:2016qyl} and \cite{Finkbeiner:2011dx} and refer the reader to those papers for details. For convenience, we summarize the main points here.

Our basis models decay with a lifetime longer than the age of the universe and each correspond to injection with a single species---either photons or $e^+ e^-$ pairs---at a single defined energy, $E_i$, which is the total kinetic energy of both injected particles. 
In other words, the particles resulting from the decay each carry an energy of $E_i/2$, which for photons is equal to $m_\chi/2$ and for electrons is $(m_\chi - 2 m_e) / 2$.
We hold the fraction of DM particles that decay equal to 1 for all basis models. The basis models are normalized by holding $p_\mathrm{ref} = 1/ \tau$ constant, where $\tau$ is the lifetime of the decaying DM in units of seconds. Each model's effect on the CMB is then fully characterized by its $f_c(z)$ functions.

A generic DM model that decays and injects energy into the universe can be approximated as a weighted sum over these basis models. Denoting the basis models as $M_i$, we can write an arbitrary model as $M = \sum_{i=1}^N \alpha_i M_i$, and the perturbation to CMB anisotropies can be written as $(\Delta C_\ell)_M = \sum_{i=1}^N \alpha_i (\Delta C_\ell)_{M_i}$. The coefficients $\alpha_i$ can be determined by the spectrum of decay products and the decay lifetime of the model $M$:
\begin{equation}
    \alpha_i = \frac{1}{p_\mathrm{ref}} \frac{1}{\tau} \frac{E_i}{m_\chi} \frac{dN_{e^+e^-, \gamma \gamma}}{d \,\mathrm{ln} E_i} d \,\mathrm{ln} E_i.
\end{equation}
Here, $\frac{dN_{e^+e^-, \gamma \gamma}}{d \,\mathrm{ln} E_i}$ describes the spectrum of $e^+ e^-$ or $\gamma \gamma$ pairs with total kinetic energy $E_i$, and $d \,\mathrm{ln} E_i$ is the spacing between the sample energies. 
For specific particle DM models, the full spectrum may be close to a delta function and so may align well with one of the basis models, but broad emission spectra that can be decomposed by the procedure described above are also possible (e.g.~in the context of decaying dark dimension gravitons~\cite{Gonzalo:2022jac} or slowly-annihilating axion quark nuggets~\cite{Majidi:2024mty}, although these specific models also have non-trivial redshift dependence that would need to be accounted for in a separate analysis).

We perform one analysis for $e^+e^-$ and photon pairs with $E_i > 27.2$ eV, and a separate analysis for photons with $E_i < 27.2$ eV (corresponding to an injection of two equal-energy photons, both below the ionization threshold). For each analysis, we generate $N$ basis models corresponding to different energy of injection and, in the first analysis, species of the injected particles. Using \texttt{ExoCLASS} with either $f_c(z)$ functions or ionization histories produced by \texttt{DarkHistory} as described in the previous section, we find the change in the $TT$, $TE$, and $EE$ CMB anisotropy power spectra and denote them $(\Delta C_\ell)_i$ for $i = 1 \dots N$. We vary $p_\mathrm{ref} = 1 / \tau$, where $\tau$ is the decay lifetime, and repeat the process. From this we can extract the derivative $\partial (\Delta C_\ell)_i / \partial p_\mathrm{ref}$ for each $i$ and $\ell$ from a polynomial fit, which allows us to test the degree of nonlinearity. We assume that the perturbations in $(\Delta C_\ell)_i$ are linear with respect to $p_\mathrm{ref}$, and this assumption holds for small energy injections. For energy injections up to the 2$\sigma$ constraint that we find later, the nonlinearity is checked to be within 10\%.

From these derivatives, we construct the $n_\ell \times N$ transfer matrix $T$ that maps energy injections into CMB anisotropy spectrum perturbations (where $n_\ell$ denotes the number of multipoles we include). The matrix $T$ has components 
\begin{equation}
    T_{\ell i} = \left( \frac{\partial (\Delta C^{TT}_\ell)_i}{\partial p_\mathrm{ref}}, \frac{\partial (\Delta C^{EE}_\ell)_i}{\partial p_\mathrm{ref}}, \frac{\partial (\Delta C^{TE}_\ell)_i}{\partial p_\mathrm{ref}} \right),
    \label{eqn:Tli}
\end{equation}
so each component $T_{\ell i}$ is a three-component vector holding the perturbations to the $TT$, $EE$, and $TE$ spectra at a certain $\ell$ for some model labeled by $i$. With this transfer matrix, we construct the $N \times N$ Fisher matrix $F_e$ as follows:
\begin{equation}
    (F_e)_{ij} = \sum\limits_{\ell} T_{\ell i}^T \cdot \Sigma_\mathrm{cov}^{-1} \cdot T_{\ell j}. 
\end{equation}
Here, $\Sigma_\mathrm{cov}$ is the covariance matrix (see e.g. Refs.~\cite{Tegmark:1996bz,Jungman:1995bz,Verde:2009tu}):
\begin{multline}
    \Sigma_\mathrm{cov} = \frac{2}{2\ell + 1} \times \\
    \begin{pmatrix}
        (C_\ell^{TT})^2 & (C_\ell^{TE})^2 & (C_\ell^{TT} C_\ell^{TE}) \\
        (C_\ell^{TE})^2 & (C_\ell^{EE})^2 & (C_\ell^{EE} C_\ell^{TE}) \\
        (C_\ell^{TT} C_\ell^{TE}) & (C_\ell^{EE} C_\ell^{TE}) & (C_\ell^{TE})^2 + C_\ell^{TT} C_\ell^{EE} \\
    \end{pmatrix}.
\end{multline}
Noise must be included in this matrix for experiments that are not cosmic variance limited (CVL). This is done by replacing $C_\ell^{TT,EE} \rightarrow C_\ell^{TT,EE} + N_\ell^{TT,EE}$, where \begin{equation}
    N_\ell = (\omega_p)^{-1} e^{\ell(\ell+1) \theta^2}.
\end{equation} Here, $N_\ell$ is the effective noise power spectrum, $\theta$ is the beam width (FWHM = $\theta \sqrt{8 \ln 2}$), and $(\omega_p)^{-1} = (\Delta T \times \mathrm{FWHM}) ^2$ is the raw beam sensitivity. We also include the effect of partial sky coverage by dividing $\Sigma_\mathrm{cov}$ by $f_\mathrm{sky} = 0.65$. In this work, we consider \textit{Planck}~\cite{Planck:2018vyg} and an experiment that is CVL in temperature and polarization up to $\ell = 2500$. We use the same noise spectrum as in Ref.~\cite{Finkbeiner:2011dx}.

Since the standard cosmological parameters and energy deposition have degenerate effects on CMB anisotropies, we also must marginalize over the cosmological parameters. We use the following set of parameters: the physical baryon density $\omega_b = \Omega_b h^2$, the physical CDM density $\omega_c = \Omega_c h^2$, the
primordial scalar spectral index $n_s$, the normalization $A_s$, the redshift of reionization $z_\mathrm{re}$, and the Hubble parameter $H_0$. We follow the same procedure described for energy deposition models to find the effect of varying the cosmological parameters on $\Delta C_\ell$, and construct their transfer matrix and Fisher matrix $F_c$.

The full Fisher matrix is then given by \begin{equation}
    F_0 = \begin{pmatrix}
        F_e & F_v \\
        F_v^T & F_c
    \end{pmatrix},
\end{equation}
where $F_e$ and $F_c$ are respectively the Fisher matrices for energy deposition and cosmological parameters, and $F_v$ contains cross terms: \begin{equation}
    (F_v)_{ij} = \sum\limits_{\ell} (T_e)_{\ell i}^T \cdot \Sigma_\mathrm{cov}^{-1} \cdot (T_c)_{\ell j}, 
\end{equation} where $T_c$ denotes the transfer matrix for changes to the cosmological parameters, and $T_e$ is the transfer matrix for energy injection. The marginalized Fisher matrix can be obtained with the block-matrix inversion and is given by $F = F_e - F_vF_c^{-1}F_v^T$. 

We diagonalize the marginalized Fisher matrix
\begin{equation}
    F = W^T \Lambda W, \,\, \Lambda = \mathrm{diag} (\lambda_1, \lambda_2, \dots, \lambda_N)
\end{equation} to get a basis of principal components, which are the eigenvectors of $F$. The $i$-th row of $W$ is the eigenvector $\vec e_i$ with eigenvalue $\lambda_i$. We normalize the eigenvectors such that they are orthonormal and order them by decreasing eigenvalue, so $\lambda_1$ is the largest eigenvalue. The principal components correspond to a set of coefficients for the basis models. 

We can now use the principal components to determine the impact of an arbitrary DM decay model on the CMB anisotropies. As previously, we write the DM model as $M = \sum_{i=1}^N \alpha_i M_i$ where $M_i$ denotes the basis models and $\vec \alpha = \{\alpha_i\}$ is a set of coefficients. We can estimate the $\Delta \chi^2$ of $M$ to be $\Delta \chi^2 = \sum_{i=1}^{N_\mathrm{max}} (\vec \alpha \cdot \vec e_i)^2 p_\mathrm{ref}^2 \lambda_i$ assuming the truth is the null hypothesis of no energy injection, with $N_\mathrm{max}$ being the number of principal components to include.

We can then constrain the decay lifetime. The 2$\sigma$ constraint corresponds to 
\begin{equation}
    p_\mathrm{ref} < \frac{2}{\sqrt{\sum_{i=1}^{N_\mathrm{max}} (\vec \alpha \cdot \vec e_i)^2 \lambda_i}}.
\end{equation}
For the basis models $M_j$ with $\alpha_i = \delta_{ij}$, this corresponds to a constraint of
\begin{equation}
    p_\mathrm{ref} < \frac{2}{\sqrt{\sum_{i=1}^{N_\mathrm{max}} (\vec e_i)_j^2 \lambda_i}} < \frac{2}{(\vec e_1)_j \sqrt{\lambda_1}}
\end{equation}
where in the second inequality we have set $N_\mathrm{max} = 1$, a good approximation when the first PC dominates. 

%%%%%%%%%%%%%%%%%%%%%%%%%%%%%%%%%%%%%%%%%%%%%%%%%%%%%%%%%%%%%%%%%
\subsection{Constraints above ionization threshold}
%%%%%%%%%%%%%%%%%%%%%%%%%%%%%%%%%%%%%%%%%%%%%%%%%%%%%%%%%%%%%%%%%

\begin{figure}
    \centering
    \includegraphics[width=\columnwidth]{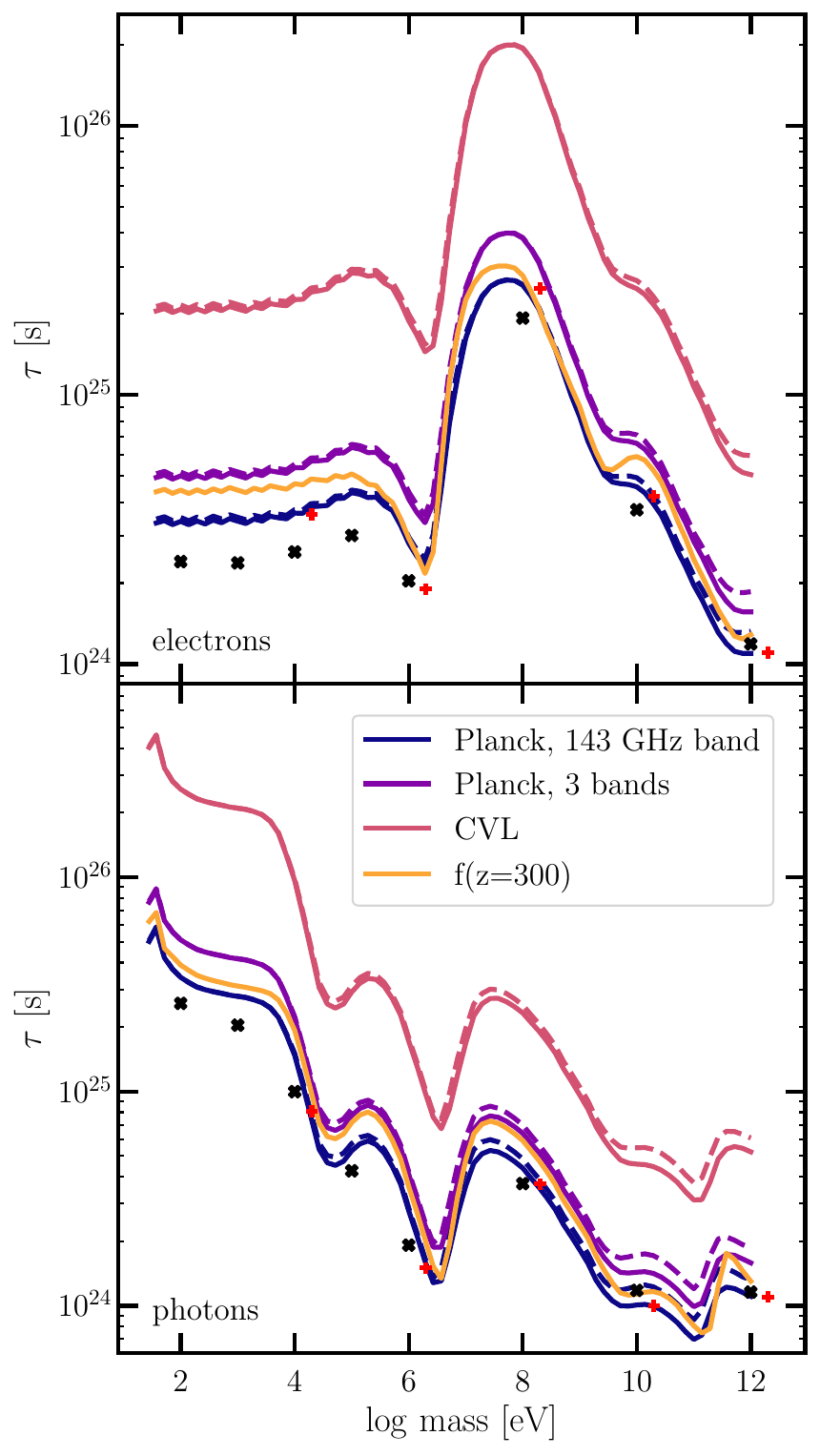}
    \caption{
        Constraints on DM decay lifetimes for an experiment that is CVL, \textit{Planck} with noise from only the 143 GHz band, and \textit{Planck} with noise from the 100, 143, and 217 GHz bands. The solid lines represent the constraint obtained from the first PC, while the dashed lines are the constraint obtained from all PCs. Black crosses show MCMC constraints, and red crosses mark the PC1 constraints from Ref.~\cite{Slatyer:2016qyl} to serve as a comparison. We also include a line for $f_{c=\mathrm{ion}}(z=300)$ (normalized arbitrarily). The $x$-axis of the electron plot shows log($m_\chi - 2 m_e$).
    }
    \label{fig:photelecconstraints}
\end{figure}

\begin{figure}
    \centering
    \includegraphics[width=\columnwidth]{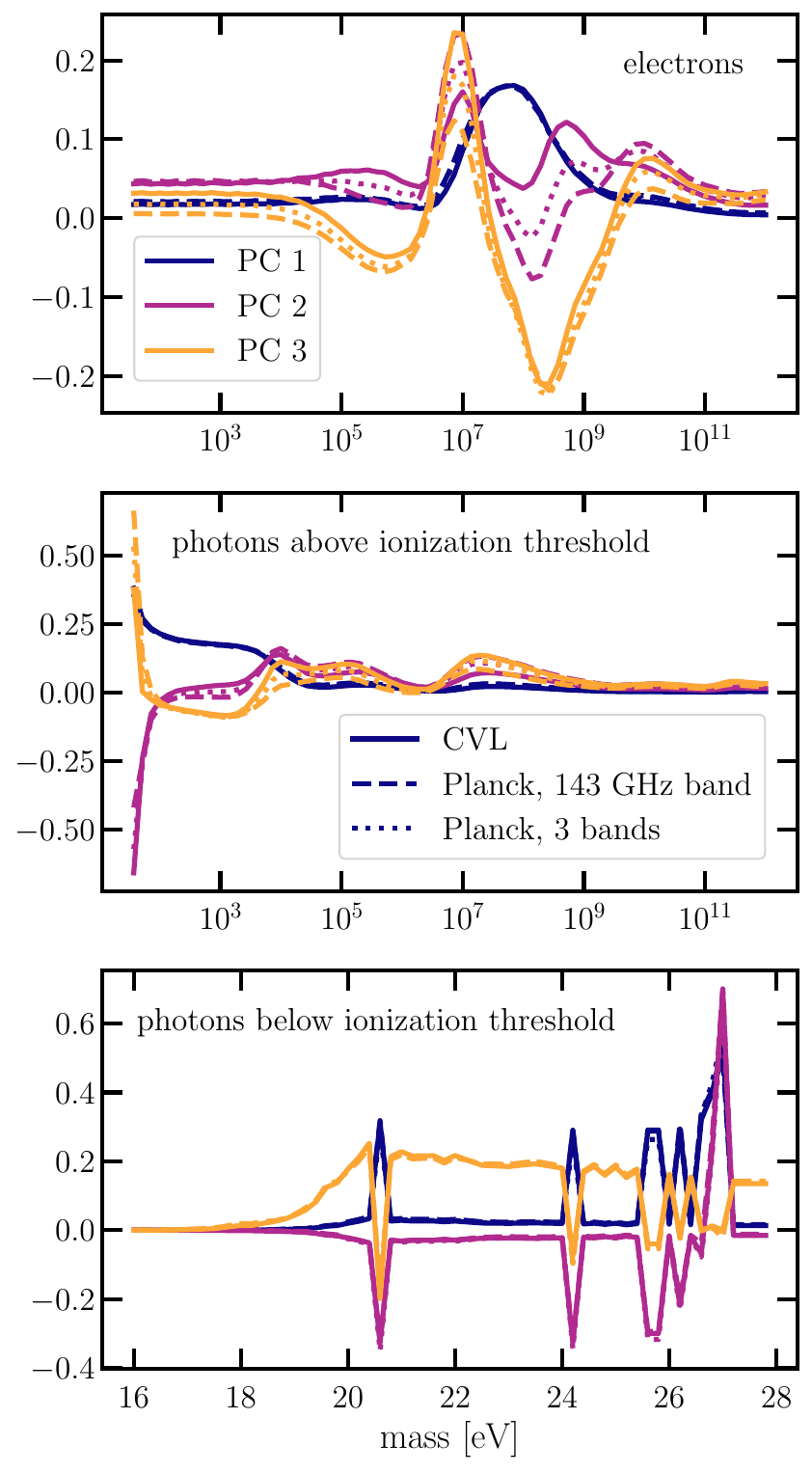}
    \caption{
        The first three PCs for a CVL experiment and \textit{Planck}. The $x$-axis of the electron plot shows log($m_\chi - 2 m_e$).
    }
    \label{fig:pc_shapes}
\end{figure}

In Fig.~\ref{fig:photelecconstraints}, we show the 2$\sigma$ constraints on the lifetime for decay into photons and electrons down to $\lesssim 100$ eV for a CVL experiment, \textit{Planck} with only the 143 GHz band, and \textit{Planck} considering the 100, 143, and 217 GHz bands (we assume the excluded bands would be used solely to remove systematics).  The solid lines represent the constraint from the first PC, while the dashed lines show the constraint from considering all PCs. We used $N=148$ basis models, evenly divided between decay into photons and electrons. For all experiments, the solid and dashed lines show excellent agreement. This is due to the dominance of the first eigenvalue; the first few eigenvalues are shown in Table~\ref{tab:pc_evals}, and the first one is two orders of magnitude greater than the second. The first PC therefore captures $\sim 98\%$ of the variance, and restricting ourselves to it gives an estimate accurate at the level of $\mathcal{O} (10 \%)$. For reference, we show the shape of the first three PCs in Fig.~\ref{fig:pc_shapes}; they remain largely the same between experiments.

\begin{table}[]
    \centering
    \begin{tabular}{c|c c c}
         Experiment & $\lambda_1$ & $\lambda_2$ & $\lambda_3$ \\
         \hline
         CVL & $5.7 \times 10^4$ & $5.5\times 10^2$ & $1.8\times10^2$\\
         Planck, 3 bands & $2.3\times 10^3$ & 28 & 16 \\
         Planck, 143 GHz & $1.0\times 10^3$ & 14 & 7.6
    \end{tabular}
    \caption{The eigenvalues corresponding to the first three principal components / eigenvectors obtained from the Fisher analysis for DM decay into electrons and photons above the ionization threshold. We consider an experiment that is CVL, \textit{Planck} with the 143 GHz band, and \textit{Planck} with the 100, 143, and 217 GHz bands. The eigenvalues have units of ($10^{25}$ s)$^2$.}
    \label{tab:pc_evals}
\end{table}

Above DM masses of $\sim1$ keV, the constraints from the first PC agree with those obtained by the first PC in Ref.~\cite{Slatyer:2016qyl} (shown with red crosses in Fig.~\ref{fig:photelecconstraints}) generally within $\sim$ 10\%.
To summarize these earlier constraints, we find a peak in the lifetime limit for decay into electrons at 30 MeV. Additionally, the shape of the constraint is approximated well by the shape of the $f_c(z)$ curve for hydrogen ionization from energy injection at $z=300$. This continues to be true for sub-keV DM decay.

Below the keV scale, we see that for decay into photons, the constraint gets stronger as the DM mass decreases. We attribute this to the increased photoionization cross section near the ionization threshold. On the other hand, the constraint for DM decaying into electrons stabilizes at $\tau \sim 3 \times 10^{24}$ s (based on the MCMC analysis). At these energies, most of the energy is stored in mass rather than kinetic energy. The positron decay product will annihilate against an ambient electron, liberating this energy and creating two $\sim 0.5$ MeV photons. The constraint on these models therefore looks similar to the constraint on DM decaying into $\sim 0.5$ MeV photons, which is also at $\tau \sim 3 \times 10^{24} s$.

In Table~\ref{tab:constraints}, we give the constraints on the decay lifetime from the \textit{Planck} experiment (using only the 143 GHz band) calculated with both only the first PC and with all PCs at the 95\% confidence level. (We also include the constraints from our MCMC analysis, described in the next section). Including higher PCs generally increases the constraint by less than 10\%, though for heavier masses, it increases on the order of 20\%. 

\begin{table}[]
    \centering
    \begin{tabular}{c|c | ccc} 
         species & $E_i$ & PC 1 & All PCs & MCMC \\
         \hline
         \multirow{8}{4em}{electrons}  & 100 eV & 0.34 & 0.35 & 0.24\\
          & 1 keV & 0.34 & 0.35 & 0.24\\
          & 10 keV & 0.36 & 0.37 & 0.26\\
          & 100 keV & 0.43 & 0.44 & 0.30\\
          & 1 MeV & 0.28 & 0.30 & 0.20\\
          & 100 MeV & 2.56 & 2.57& 1.93\\
          & 10 GeV & 0.46 & 0.49 & 0.38\\
          & 1 TeV & 0.11 & 0.13 & 0.12\\
          \hline
         \multirow{8}{4em}{photons}  & 100 eV & 3.41 & 3.41 & 2.58\\
          & 1 keV & 2.78 & 2.79 & 2.04\\
          & 10 keV & 1.48 & 1.51 & 1.00\\
          & 100 keV & 0.53 & 0.57 & 0.43\\
          & 1 MeV & 0.28 & 0.29 & 0.19\\
          & 100 MeV & 0.44 & 0.48 & 0.37\\
          & 10 GeV & 0.10 & 0.12 & 0.12\\
          & 1 TeV & 0.11 & 0.13 & 0.12\\
    \end{tabular}
    \caption{Lower bounds from \textit{Planck} (noise from the 143 GHz band) on the decay lifetime in units of $10^{25}$ seconds at 95\% confidence, for a range of energies and species. The first column shows the constraint from the first PC, the second shows the constraint from all PCs, and the third shows the MCMC constraint.}
    \label{tab:constraints}
\end{table}

\subsection{Validation with MCMC and \textit{Planck} 2018 data}

Fisher analysis and PCA assume linearity and Gaussian likelihoods. These are good approximations to estimate the constraints on the DM decay lifetime, but it is useful to confirm these estimates and find the true posteriors of the cosmological parameters using Markov Chain Monte Carlo (MCMC) methods. To do so, we use the publicly available \texttt{Monte Python} code interfaced with \texttt{ExoCLASS}.

We use the \textit{Planck} 2018 data with four likelihoods: the \texttt{lowl\_TT} likelihood from the temperature spectrum over $2 \le \ell < 30$, the \texttt{lowl\_EE} likelihood from the $EE$ power spectrum over $2 \le \ell < 30$, the \texttt{highl\_TTTEEE} likelihood from the TT,TE,EE spectra at $\ell > 29$, and the lensing likelihood. We assume flat priors on the six cosmological parameters: $\omega_b$, $\omega_c$, $n_s$, $A_s$, $z_\mathrm{re}$, and $H_0$, along with a parameter describing the DM decay lifetime, $p_\mathrm{ref} = 1 / \tau$ \footnote{The choice of prior on the DM energy injection parameter can have an effect on the MCMC results; for example, choosing a flat prior on log($p_\mathrm{ref}$) generally results in a stronger constraint. We choose a flat prior on $p_\mathrm{ref}$ following Ref.~\cite{Slatyer:2016qyl}, with limits $0 < p_\mathrm{ref}/(10^{-25} s^{-1}) < 100$.
}. 

We treat energy deposition as described in previous sections, passing in the $f_c(z)$ functions from \texttt{DarkHistory} to \texttt{ExoCLASS} for decay into electrons and photons above the ionization threshold. 

\begin{figure*}
    \centering
    \includegraphics[width=\textwidth]{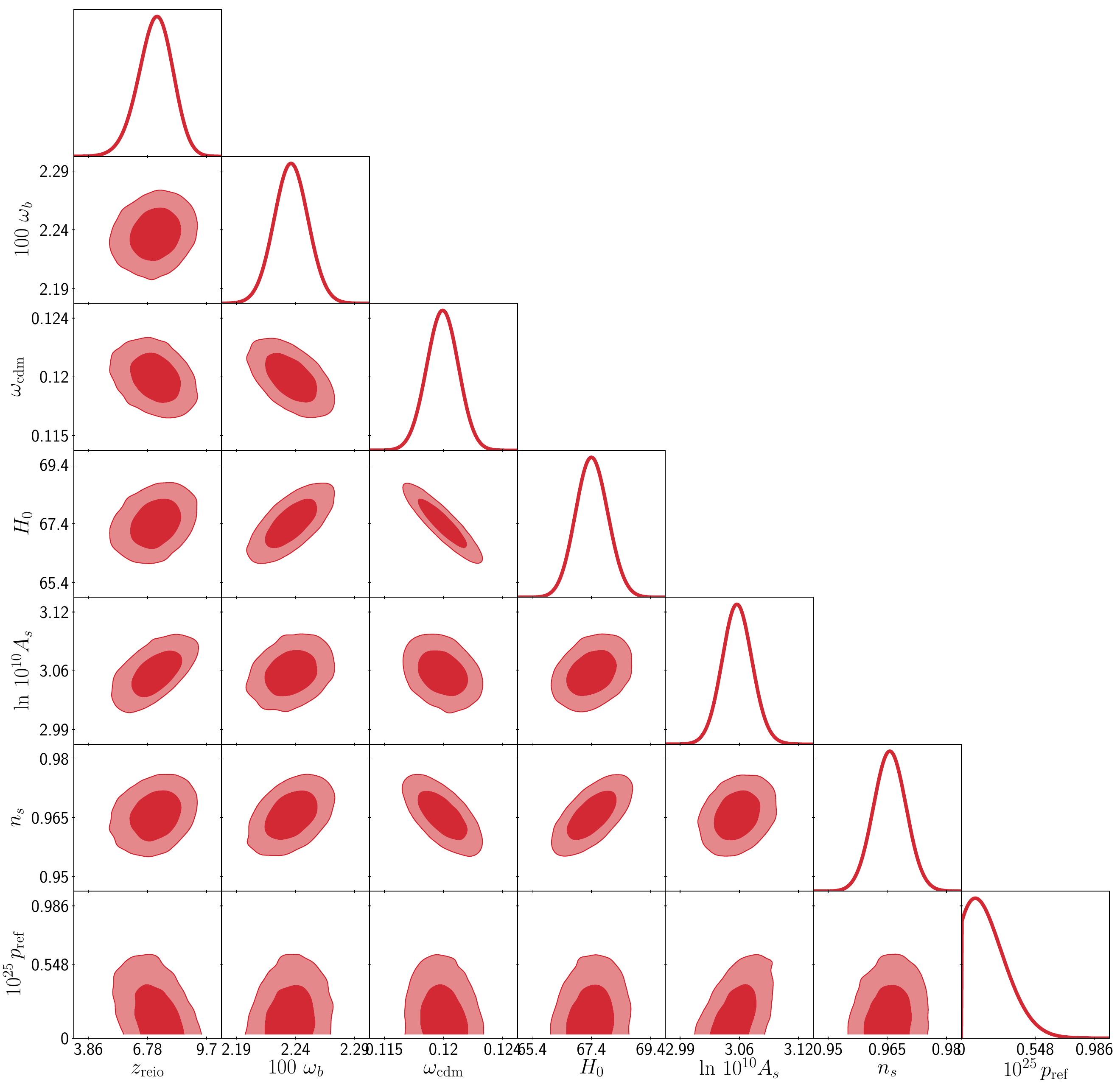}
    \caption{The marginalized 2D posterior probability distributions for the cosmological parameters and $p_\mathrm{ref} = 1/\tau$. This example assumes decay into two photons with $m_\chi = 1$ keV.}
    \label{fig:mcmctriangle}
\end{figure*}

We use the Gelman-Rubin statistic, $R$, to measure convergence, running chains until $R-1$ falls below 0.01 for each parameter. We then marginalize over the nuisance parameters to obtain constraints and contour plots; an example of the contour plots is shown in Fig.~\ref{fig:mcmctriangle}. 

In Fig.~\ref{fig:photelecconstraints}, we show our 95\% confidence level MCMC constraints for several injection energies along with constraints from PCA, and we also list the constraints in Table~\ref{tab:constraints}. Generally, the two are in good agreement with each other, though the MCMC constraints are slightly weaker as expected from the non-Gaussianity of the likelihood~\cite{Verde:2009tu}. This confirms that the PCs can be used to estimate constraints on the DM decay lifetime down to energy injection at sub-keV levels (but above the ionization threshold).

%%%%%%%%%%%%%%%%%%%%%%%%%%%%%%%%%%%%%%%%%%%%%%%%%%%%%%%%%%%%%%%%%
\subsection{Constraints below ionization threshold}
%%%%%%%%%%%%%%%%%%%%%%%%%%%%%%%%%%%%%%%%%%%%%%%%%%%%%%%%%%%%%%%%%

\begin{figure}
    \centering
    \includegraphics[width=\columnwidth]{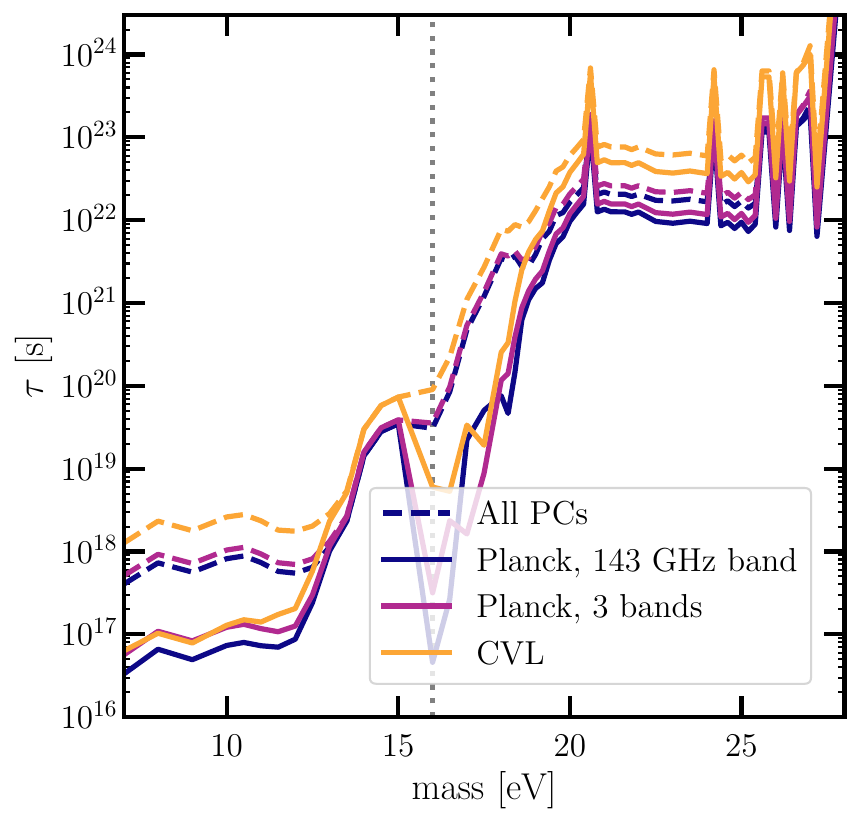}
    \caption{Constraints on the DM decay lifetime for decay into photons below the ionization threshold, obtained from PCA. The solid lines are the constraints from the first PC, and the dashed lines are the constraints with all PCs. 
    For numerical stability across this large range of lifetimes, we perform two separate PCAs above and below 16 eV, as denoted by the dotted gray line.}
    \label{fig:photlowconstraints}
\end{figure}

Fig.~\ref{fig:photlowconstraints} shows the PCA 2$\sigma$ constraints on the lifetime for decay into two photons with $m_\chi < 27.2$ eV. 
We use the same experiments as in the previous section and again show the constraints from both the first PC and from all PCs. 
We perform separate PCAs above ($N=48$) and below ($N=13$) $m_\chi = 16$ eV: since the effect of decays on the thermal history drops sharply towards lower DM masses (see Fig.~\ref{fig:delta_xe_lowmass}), we must perform the PCA with a higher value of $p_\mathrm{ref}$ in order for the derivatives in Eqn.~\ref{eqn:Tli} to be numerically stable. 
The first three eigenvalues are given in Table~\ref{tab:pc_evals_low}, where we see that the first eigenvalue dominates by a little less than an order of magnitude. Higher PCs are therefore more important in this energy range. 
We only show constraints down to masses of about 7 eV as below this value, the results are not numerically stable.
However, below this value, the constraint will likely die away rapidly, since photons injected with energies less than 3.4 eV will need to find higher excited states (principal quantum number $n \ge 3$) to ionize, and the abundance of such states is highly suppressed.

\begin{table}[]
    \centering
    \begin{tabular}{c|c|c c c}
         & Experiment & $\lambda_1$ & $\lambda_2$ & $\lambda_3$ \\
         \hline
         \multirow{3}{3em}{above 16 eV} & CVL & $1.4 \times 10^{-1}$ & $4.9 \times 10^{-2}$ & $2.0 \times 10^{-3}$ \\
         & Planck, 3 bands & $1.2 \times 10^{-2}$ & $3.1 \times 10^{-3}$ & $3.1 \times 10^{-4}$ \\
         & Planck, 143 GHz & $6.3 \times 10^{-3}$ & $1.6 \times 10^{-3}$ & $2.1 \times 10^{-4}$ \\
         \hline
         \multirow{3}{3em}{below 16 eV} & CVL & $3.9 \times 10^{-10}$ & $1.8 \times 10^{-12}$ & $2.6 \times 10^{-13}$ \\
         & Planck, 3 bands & $1.1 \times 10^{-10}$ & $2.8 \times 10^{-13}$ & $3.5 \times 10^{-14}$ \\
         & Planck, 143 GHz & $8.9 \times 10^{-11}$ & $1.8 \times 10^{-13}$ & $2.7 \times 10^{-14}$
    \end{tabular}
    \caption{The eigenvalues corresponding to the first three principal components / eigenvectors obtained from the Fisher analysis for DM decay into photons below the ionization threshold. The eigenvalues have units of ($10^{25}$ s)$^2$.}
    \label{tab:pc_evals_low}
\end{table}

Compared to the constraints at higher energies, these constraints are a few orders of magnitudes weaker, excluding lifetimes around the level of $10^{22}$ seconds. The shape is also no longer correlated with the energy deposition function in the channel of hydrogen ionization at $z=300$; $f_{\text{c=ion}}(z) = 0$ in this energy range since the photons are not able to directly ionize hydrogen. Instead, comparing Figs.~\ref{fig:photlowconstraints} and \ref{fig:delta_xe_lowmass}, the shape is approximated by the difference in the free electron fraction at redshifts around recombination, with resonances near energies corresponding to hydrogen spectral lines. 
Around 12 eV, the lifetime constraint falls to $\sim 10^{18}$ s, which is comparable to the age of universe. 
Consequently, at this point we begin to lose sensitivity to scenarios where 100\% of the DM is decaying (or equivalently, our constraints are overtaken by gravitational bounds on the DM decay lifetime, e.g.~\cite{Poulin:2016nat,Simon:2022ftd}), since the DM must have a lifetime appreciably longer than the age of the universe in this case. 
There could still be a non-trivial constraint on a subdominant fraction ($\sim 10^{-5}$ or larger) of the DM decaying with a lifetime as short as $\sim 10^{13}$ s (the approximate age of the universe during the epoch of last scattering).

\begin{figure}
    \centering
    \includegraphics[width=\columnwidth]{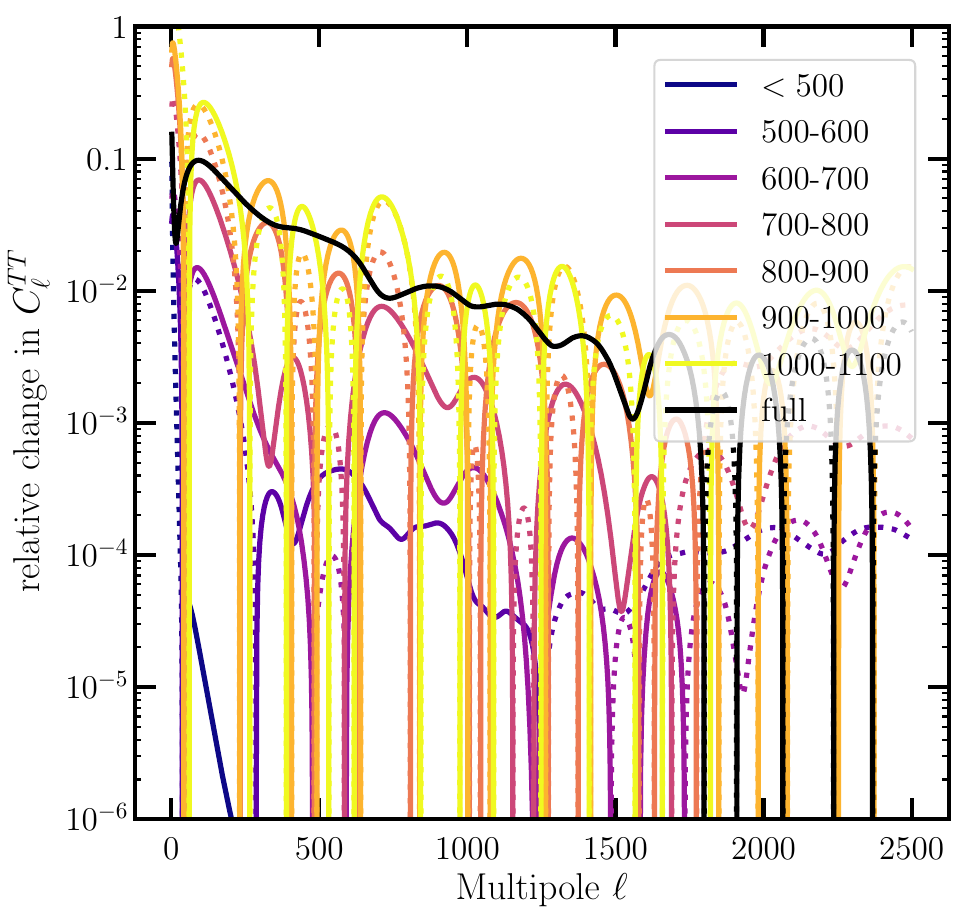}
    \caption{The relative change in $C_\ell^{TT}$ (after marginalizing over the cosmological parameters) when including the effects of energy injection on the ionization history only at specific redshift ranges, relative to the case with no energy injection. The energy injection model used is DM decaying into photons with lifetime $5 \times 10^{21}$ s at a mass of 22 eV; the black line shows the relative change in $C_\ell^{TT}$ with the effects of energy injection included at all redshifts. Dotted lines show the negative of the plotted values.}
    \label{fig:rel_TT_change}
\end{figure}

As previously discussed in Section~\ref{sec:eng_inj}, we expect redshifts around recombination at $z \sim 1000$ to dominate the effect on the power spectrum. To verify this, we plot the contribution to $\Delta C_\ell$ of different redshift ranges in Fig.~\ref{fig:rel_TT_change}. It is clear that redshifts around recombination have the greatest effect on the perturbations to the $C_\ell$'s. The range $900 < z < 1000$ contributes the most to the change in $C_\ell$, and the contribution decreases with decreasing redshift. 

In Table~\ref{tab:constraints_low}, we give the constraints on the lifetime for decay into photons at several energies below the ionization threshold. Away from the peaks, taking into account all PCs nearly doubles the constraint on the lifetime compared to only considering PC 1. Near the peaks, including all PCs increases the constraint by around 20\%. On the other hand, higher PCs become much more important at masses below 20 eV, with a nearly 500\% increase in the constraint when taking them into account at a mass of 18 eV.

We did not perform a MCMC analysis for models that decay into photons below the ionization threshold, as this would require running \texttt{DarkHistory} for each step in order to get an ionization history to feed into \texttt{ExoCLASS}, which would take several hours per step; we leave it to future work to verify the PCA estimates on the DM decay lifetime in this range. However, given the success of the PCA in the higher mass case, we expect our PCA here to also give an accurate result.

\begin{table}[]
    \centering
    \begin{tabular}{c|c | c c c}
         species & $E_i$ & PC 1 & All PCs \\
         \hline
         \multirow{7}{4em}{photons} & 10 eV & $7.3 \times 10^{-6}$ & $8.2 \times 10^{-5}$ \\ 
         & 18 eV & 0.0076 & 0.35 \\
          & 20 eV & 0.97 & 1.6 \\
          & 20.4 eV & 12.7 & 14.3 \\
          & 22 eV & 1.2 & 2.1 \\
          & 24.2 eV & 11.5 & 13.2 \\
          & 27 eV & 20.0 & 24.5 \\
    \end{tabular}
    \caption{Lower bounds from \textit{Planck} (143 GHz band) on the lifetime for decay into photons in units of $10^{22}$ seconds at 95\% confidence, for a range of energies. The first column shows the constraint from the first PC, and the second shows the constraint from all PCs.}
    \label{tab:constraints_low}
\end{table}

%%%%%%%%%%%%%%%%%%%%%%%%%%%%%%%%%%%%%%%%%%%%%%%%%%%%%%%%%%%%%%%%%
%%%%%%%%%%%%%%%%%%%%%%%%%%%%%%%%%%%%%%%%%%%%%%%%%%%%%%%%%%%%%%%%%
\section{Summary and comparison to existing constraints}
\label{sec:constraints}
%%%%%%%%%%%%%%%%%%%%%%%%%%%%%%%%%%%%%%%%%%%%%%%%%%%%%%%%%%%%%%%%%
%%%%%%%%%%%%%%%%%%%%%%%%%%%%%%%%%%%%%%%%%%%%%%%%%%%%%%%%%%%%%%%%%

\begin{figure*}
    \centering
    \includegraphics[width=\textwidth]{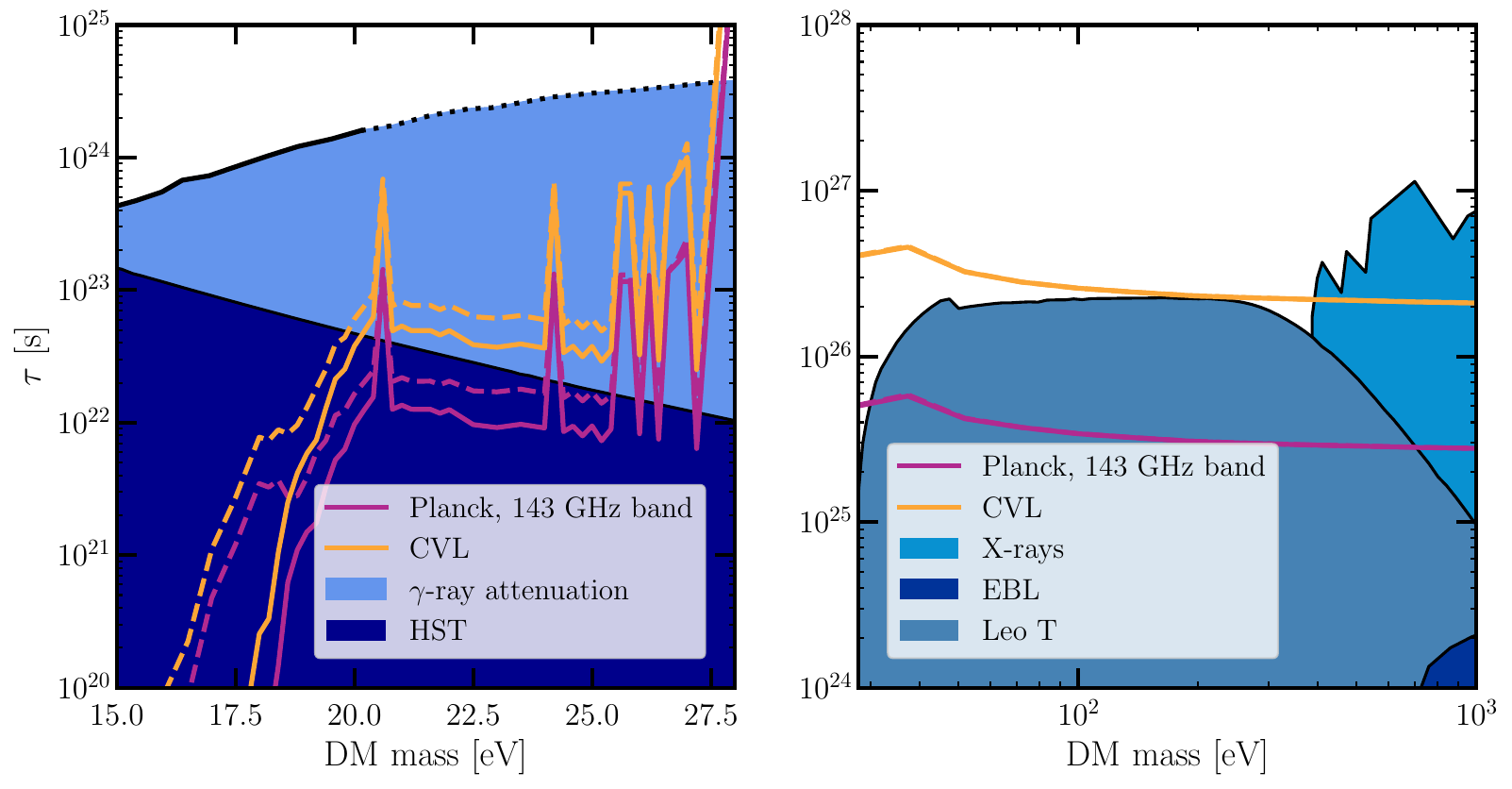}
    \caption{Comparison of our PCA lower bounds to existing constraints of the lifetime of DM decaying into photons.
    We include results using limits on X-ray backgrounds and extragalactic background light (EBL) from Ref.~\cite{Cadamuro:2011fd}, optical background observations from the Hubble Space Telescope~\cite{Carenza:2023qxh}, gamma-ray attenuation bounds (the dotted line indicates where absorption by neutral hydrogen has been neglected)~\cite{Bernal:2022xyi}, and anomalous heating of the Leo T dwarf galaxy~\cite{Wadekar:2021qae}. Dashed lines indicate the constraint obtained from all PCs, while the solid line shows the constraint from the first PC. The plot on the right shows both constraints, but they are indistinguishable at the level of the plot.
    }
    \label{fig:compare}
\end{figure*}

We can now ask how our constraints compare to existing bounds on the DM decay lifetime. For decay into electrons, and for decay into photons at masses above the keV scale, our constraints simply update earlier work~\cite{Slatyer:2016qyl}. In general, indirect searches have placed stronger constraints on the lifetime for these cases, on the order of $10^{28}$ seconds for monochromatic photons and $10^{26}$ seconds for $e^+e^-$ pairs (see e.g. Refs.~\cite{Cirelli:2023tnx,Sun:2023acy}), although the CMB constraints avoid many of the systematics associated with indirect searches and are quite insensitive to the spectrum of final-state particles. 
Our constraints on decay into photons at sub-keV DM masses, however, are within an order of magnitude of previous bounds such as the one from anomalous heating of the Leo T dwarf galaxy~\cite{Wadekar:2021qae}; Fig.~\ref{fig:compare} shows a comparison to existing bounds. 
The lower bound from CMB anisotropies also has the potential to become the strongest for decaying DM in the 100 eV mass range as experiments improve towards being CVL; looking at the eigenvalues listed in Table~\ref{tab:pc_evals}, we expect this bound to improve by a factor of 5. 

For DM decaying into photons below the ionization threshold, our constraint is competitive with that from optical background observations by the Hubble Space Telescope (HST)~\cite{Carenza:2023qxh} above masses of 20.4 eV. Above 20.4 eV, the constraint from gamma-ray attenuation is 1-2 orders of magnitude stronger, but does not take into account absorption of photons by neutral hydrogen~\cite{Bernal:2022xyi}. Below this mass, our constraint falls off. 
We also expect this bound to improve by a factor of 2 to 3 in decay lifetime for an experiment that is CVL up to $\ell=2500$, based on the eigenvalues in Table~\ref{tab:pc_evals_low}. We note that Ref.~\cite{Bolliet:2020ofj} performed a similar analysis around this mass range with data from COBE/FIRAS, EDGES, and \textit{Planck}, but treated power into ionization in a simplified way as they were not focused on this energy region. 

In general, our constraints are competitive with other bounds on the lifetime, especially in the mass range between 20.4 eV and $\sim$ 100 eV. 
Additionally, our CMB constraints avoid some of the systematic uncertainties that accompany previous bounds (such as astrophysical uncertainties regarding the modeling of DM haloes or EBL), providing a complementary result.

%%%%%%%%%%%%%%%%%%%%%%%%%%%%%%%%%%%%%%%%%%%%%%%%%%%%%%%%%%%%%%%%%
%%%%%%%%%%%%%%%%%%%%%%%%%%%%%%%%%%%%%%%%%%%%%%%%%%%%%%%%%%%%%%%%%
\section{Conclusion}
\label{sec:conclusion}
%%%%%%%%%%%%%%%%%%%%%%%%%%%%%%%%%%%%%%%%%%%%%%%%%%%%%%%%%%%%%%%%%
%%%%%%%%%%%%%%%%%%%%%%%%%%%%%%%%%%%%%%%%%%%%%%%%%%%%%%%%%%%%%%%%%

We have extended the bounds on the lifetime of decaying DM down to sub-keV masses, using principal component analysis. This analysis was divided between decay products that can deposit energy directly into the channel of hydrogen ionization, and products that can only directly excite hydrogen, i.e. photons below the ionization threshold. In both cases, and especially for the prior case, we find that the effect on the CMB anisotropies can be approximately captured by a single parameter, although the parameter for each case is different. The effect of generic DM decay models on CMB anisotropies can then be found with a simple weighted sum. 
We have verified our PCA analysis with a MCMC using \textit{Planck} 2018 data. 

For sub-keV DM decaying into photons above the ionization threshold, we find a lower bound on the lifetime on the order of $10^{26}$ seconds, while for sub-keV DM decaying into electrons the constraint is on the order of $10^{25}$ seconds. The shape of the constraint in both cases is approximated well by the energy deposition function $f_{c=\mathrm{ion}}(z=300)$ into the channel of hydrogen ionization. The current constraint is also expected to increase significantly as experiments improve; a CVL experiment up to $\ell = 2500$ in temperature and polarization could thus potentially place the most stringent existing constraint on the lifetime of decay into photons around the 100 eV mass range. 

For DM decaying into photons below the ionization threshold, we find lifetime constraints mostly on the order of $10^{22}$ seconds, with peaks around masses corresponding to hydrogen spectral lines having a stronger lifetime bound of order $10^{23}$ seconds. Rather than its shape being determined by $f_{c=\mathrm{ion}}(z=300)$, the constraint is instead sensitive to the change in the free electron fraction $x_e$ at redshifts around recombination. Our constraints are competitive with previous bounds at a mass above 20.4 eV and thus provide a complementary result; additionally, as experiments improve, the CMB constraint has the potential to improve by a factor of $\sim$3.

%%%%%%%%%%%%%%%%%%%%%%%%%%%%%%%%%%%%%%%%%%%%%%%%%%%%%%%%%%%%%%%%%
%%%%%%%%%%%%%%%%%%%%%%%%%%%%%%%%%%%%%%%%%%%%%%%%%%%%%%%%%%%%%%%%%
\section*{Acknowledgements}
%%%%%%%%%%%%%%%%%%%%%%%%%%%%%%%%%%%%%%%%%%%%%%%%%%%%%%%%%%%%%%%%%
%%%%%%%%%%%%%%%%%%%%%%%%%%%%%%%%%%%%%%%%%%%%%%%%%%%%%%%%%%%%%%%%%

C.X. was supported by the MIT Undergraduate Research Opportunities Program, specifically the Ralph L. Evans (1948) Endowment Fund.
W.Q. was supported by the National Science Foundation Graduate Research Fellowship under Grant No. 2141064.
T.R.S. was supported by the Simons Foundation (Grant Number 929255, T.R.S) and by a Radcliffe Fellowship and Guggenheim Fellowship. T.R.S. thanks the Kavli Institute for Theoretical Physics (KITP) and the Aspen Center for Physics for their hospitality; this research was supported in part by grant no.~NSF PHY-2309135 to KITP, and performed in part at the Aspen Center for Physics, which is supported by National Science Foundation grant PHY-2210452. This work was supported by the U.S. Department of Energy, Office of Science, Office of High Energy Physics of U.S. Department of Energy under grant Contract Number DE-SC0012567. 

This work made use of 
\texttt{NumPy}~\cite{Harris:2020xlr}, 
\texttt{SciPy}~\cite{2020NatMe..17..261V}, 
\texttt{Jupyter}~\cite{Kluyver2016JupyterN}, 
\texttt{matplotlib}~\cite{Hunter:2007ouj}, 
and \texttt{tqdm}~\cite{daCosta-Luis2019},
as well as Webplotdigitizer~\cite{Rohatgi2022}.

\bibliography{refs}
\end{document}